\documentclass[secnumarabic,pra,aps,12pt,reprint,floatfix,nofootinbib,
tightenlines,nobibnotes]{revtex4-1}

\usepackage{hyperref}
\usepackage{graphicx}
\usepackage{soul}
\usepackage{xcolor}
\usepackage{epstopdf}
\usepackage[caption=false]{subfig}
\usepackage{multirow}
\usepackage{amsmath}
\usepackage{amsfonts}
\usepackage{color}
\usepackage{url}
\usepackage{appendix}
\usepackage{amssymb}

\usepackage[T1]{fontenc} 
\usepackage{microtype} 

\usepackage[english]{babel} 

\usepackage[top=2cm,bottom=4cm,left=2cm,right=2cm]{geometry} 
\usepackage{booktabs} 

\usepackage{lettrine} 

\usepackage{enumitem} 
\setlist[itemize]{noitemsep} 

\usepackage{titlesec} 
\titleformat{\section}[block]{\large\scshape\centering}{\thesection.}{1em}{} 
\titleformat{\subsection}[block]{\large}{\thesubsection.}{1em}{} 

\usepackage{fancyhdr} 
\begin{document}
\author{Huy Tuong Cao}
\affiliation{Department of Physics and The Institute of Photonics and Advanced Sensing (IPAS),\\
University of Adelaide, SA, 5005, Australia\\
OzGrav, Australian Research Council Centre of Excellence for Gravitational Wave Discovery}
\author{Daniel D. Brown}
\email{daniel.d.brown@adelaide.edu.au}
\affiliation{Department of Physics and The Institute of Photonics and Advanced Sensing (IPAS),\\
University of Adelaide, SA, 5005, Australia\\
OzGrav, Australian Research Council Centre of Excellence for Gravitational Wave Discovery}
\author{Peter Veitch}
\affiliation{Department of Physics and The Institute of Photonics and Advanced Sensing (IPAS),\\
University of Adelaide, SA, 5005, Australia\\
OzGrav, Australian Research Council Centre of Excellence for Gravitational Wave Discovery}
\author{David J. Ottaway}
\affiliation{Department of Physics and The Institute of Photonics and Advanced Sensing (IPAS),\\
University of Adelaide, SA, 5005, Australia\\
OzGrav, Australian Research Council Centre of Excellence for Gravitational Wave Discovery}
\date{\today} 
\begin{abstract}
Knowledge of the intensity and phase profiles of spectral components in a coherent optical field is critical for a wide range of high-precision optical applications. One of these is interferometric gravitational wave detectors, which rely on such fields for precise control of the experiment. Here we demonstrate a new device, an \textit{optical lock-in camera}, and highlight how they can be used within a gravitational wave interferometer to directly image fields at a higher spatial and temporal resolution than previously possible. This improvement is achieved using a Pockels cell as a fast optical switch which transforms each pixel on a sCMOS array into an optical lock-in amplifier. We demonstrate that the optical lock-in camera can image fields with 2~Mpx resolution at 10~Hz with a sensitivity of -62~dBc when averaged over 2s.
\end{abstract}


\title{An optical lock-in camera for advanced gravitational wave interferometers}

\maketitle

The detection of gravitational waves (GW)~\cite{PhysRevLett.116.131103} has ushered in a new era of gravitational and multi-messenger astronomy. Improving the sensitivity of current and next-generation detectors will ensure that they fulfill their potential to observe this exciting new window on the universe. Reaching these goals however will require a significant reduction in quantum noise which can be achieved by increasing both the circulating optical power stored within the interferometer and the use of squeezed light~\cite{kwee2004, Toyra2017}. To achieve the maximum benefit from these upgrades it is essential that precise control of the optical beams circulating within the interferometer is achieved.

\begin{figure*}[!ht]
\centering
\includegraphics[width=0.9\textwidth]{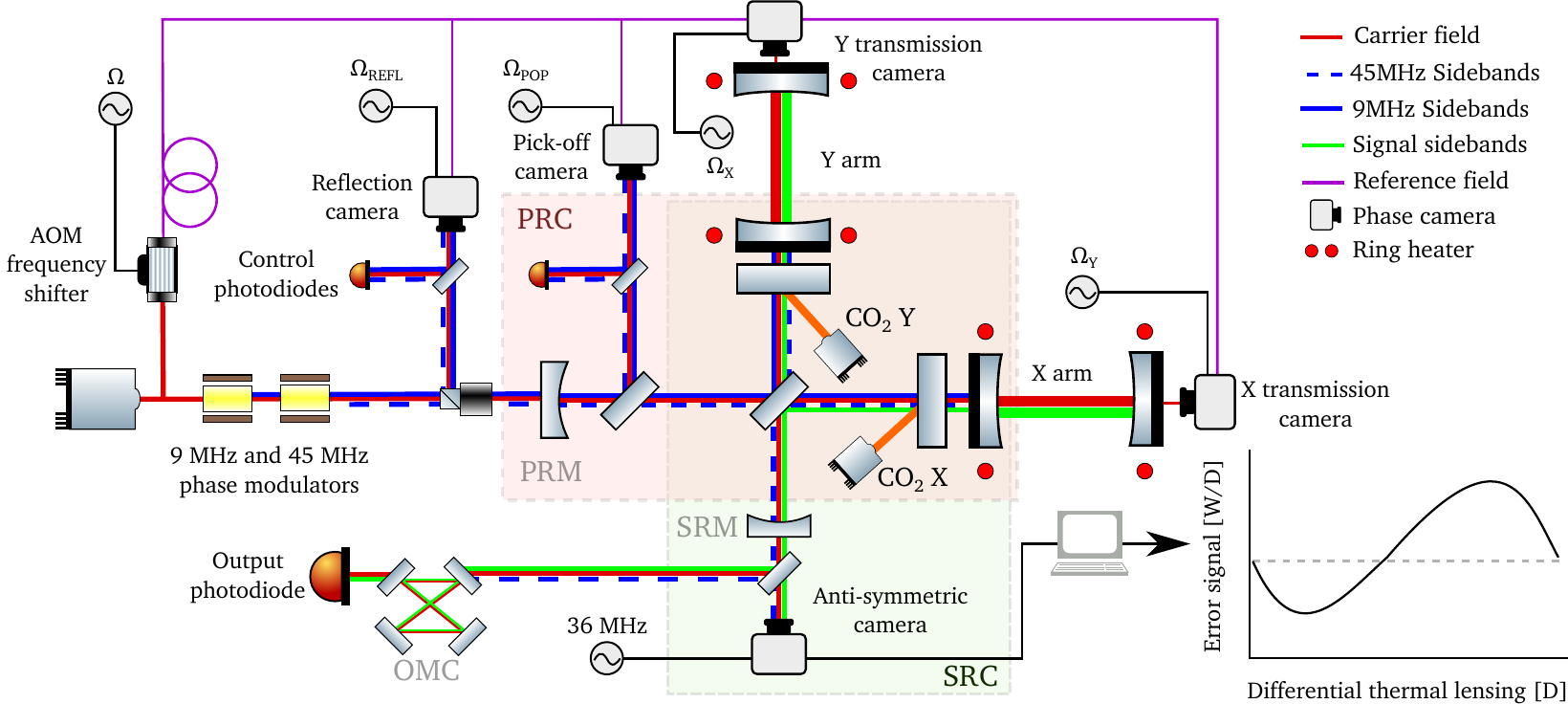}
\caption{Schematic of a detector similar to LIGO and possible locations for phase cameras. Highlighted are the power recycling (PRC), signal recycling (SRC), output mode cleaner (OMC), the arm cavities (XARM and YARM), the RF modulation sidebands used for control the interferometer, and the cavities in which they resonate. The thermal actuators used to mode-match the interferometer are also shown: ring heaters around each arm-cavity mirror and CO$_{2}$ laser beams incident on compensation plates. Five potential locations for phase cameras are shown. Combining the sampled field with a reference field that is offset-locked to the  main laser, as shown, and choosing the appropriate switching frequency would allow the amplitude of each field to be mapped.
}
\label{Fig:ifo_fields}
\end{figure*}

Optical heterodyne techniques, such as Pound-Drever-Hall locking, are used extensively throughout ground-based GW detectors to generate \textit{error signals} with which to control the relative positions and alignments of the suspended optics~\cite{Black:2001, Fritschel:98, Mueller:00, AGATSUMA2016598}. These systems use radio-frequency~(RF) phase-modulation sidebands which are imposed on a carrier field and resonate within the different optical cavities of the interferometer, as shown in Fig.~\ref{Fig:ifo_fields}. The RF beat-notes which are then demodulated at various photodiode outputs are used to produce the error signals.

Wavefront mismatches from static deformations in the optics or thermal distortions from bulk or small, highly absorbing defects can introduce significant time-dependent offsets in error signal set points. This is due to the sidebands experiencing different resonant conditions within the interferometer and becoming distorted relative to each other resulting in poor spatial overlap. This imbalance leads to a degradation of the error signals and the performance of the control systems. Thus, detailed knowledge of all the carrier and sideband fields is required to fully understand control sensing issues and design adequate solutions for enhancing the detectors further.

Thermal compensation systems (TCS) in LIGO~\cite{Brooks:16} and Virgo~\cite{Rocchi_2012} aim to reduce the effects of this absorption-induced wavefront distortion in the current generation of detectors. However, these systems use auxiliary probe beams to sense the distortion in core optics. The scanning \textit{phase camera} \cite{Goda:04, Agatsuma:19}, by contrast, uses the interferometer fields directly to investigate the effect of thermal deformations. They offer a significant potential in helping commission and operate detectors, such as more efficient and optimal tuning of TCS, but have yet to reach their full potential.

The scanning phase camera, developed by \citet{Goda:04}, measures the transverse intensity and phase distribution of specific RF frequency components within a coherent field. Other common methods for measuring wavefronts---such as Hartmann sensors~\cite{Brooks:07, Shack:71,Stoklasa:2014}, phase retrieval methods~\cite{Zhang:2017}, spatial wavefront sampling ~\cite{Soldevila:18}, holography~\cite{Schulze:12, Dudley:14}, and other interferometric techniques~\cite{Wolfling:05, TarquinRalph:17, Love:05}---measure the superposition of all spectral components in a beam and thus lack any frequency-selectivity. Phase cameras alone enable analysis of each component of the interferometer field.

Phase cameras use the heterodyne beat between the interferometer field and a reference field at a single transverse location, which is recorded using a photodiode and demodulated at the beat frequency of interest. Scanning the field over the photodiode using movable mirrors provides a 2D intensity and phase map. The maximum achievable spatial and temporal resolution is limited by the mechanical resonances of the scanner and the signal processing. Additionally, the scanning may cause mechanical vibrations and time-varying light scattering, which could be unsuitable for highly sensitive systems such as GW detectors. 

\begin{figure}[ht]
\includegraphics[width=0.9\columnwidth]{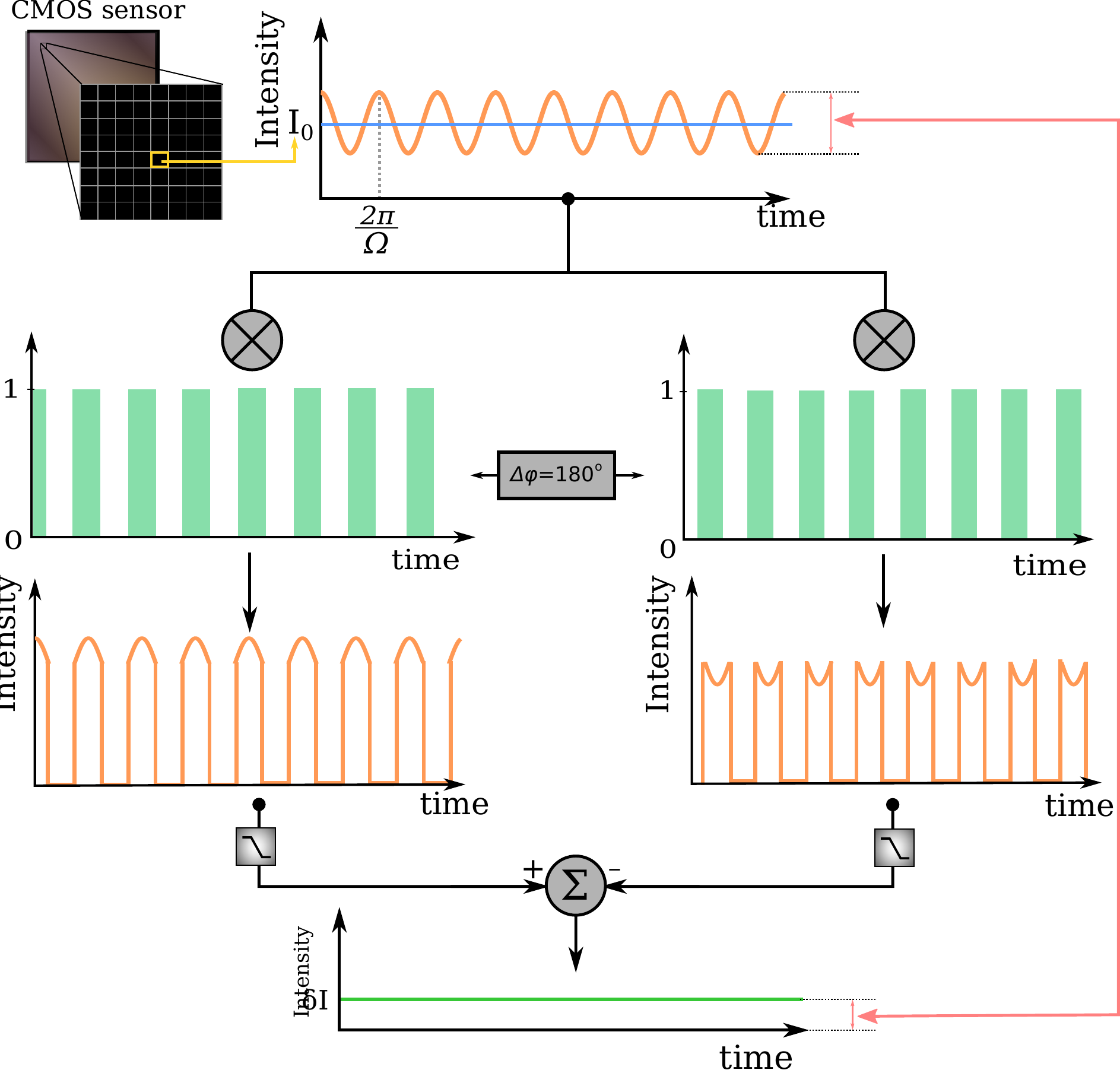}
\caption{The operation of the new camera can be visualized by considering the beat signal measured by a single pixel. Synchronous intensity modulation of the incident light field at frequency $\Omega$ allows the pixel to extract a DC signal that is a function of the magnitude and phase of the beat.}
\label{Fig:WorkingPrinciple}
\end{figure}

In this paper we describe and demonstrate an alternative phase camera approach that has no moving parts. It produces a 2-dimensional map of the intensity and phase of a spectral component within a coherent light field. Its temporal and spatial resolution is determined by the frame rate and pixel size of the camera, thus enabling high resolution and fast capture rates. This is achieved by using a Pockels cell as a fast optical switch which transforms the array of pixels into a parallel array of optical lock-in amplifiers.

 We begin with an overview of the operating principle of the optical lock-in phase camera. In Section~\ref{sec:gw} we discuss potential applications of the camera in a GW detector. The experimental realization of the phase camera is outlined in Section~\ref{sec:expt}. Measured intensity and phase maps are compared with the predictions of a numerical model of the test system. Finally, we demonstrate that the sensitivity is shot-noise limited and can thus be improved simply by averaging.

\section{Principle of Operation} \label{sec:workingPrinc}

\begin{figure*}[ht!]
	\centering
	\includegraphics[width=0.99\textwidth]{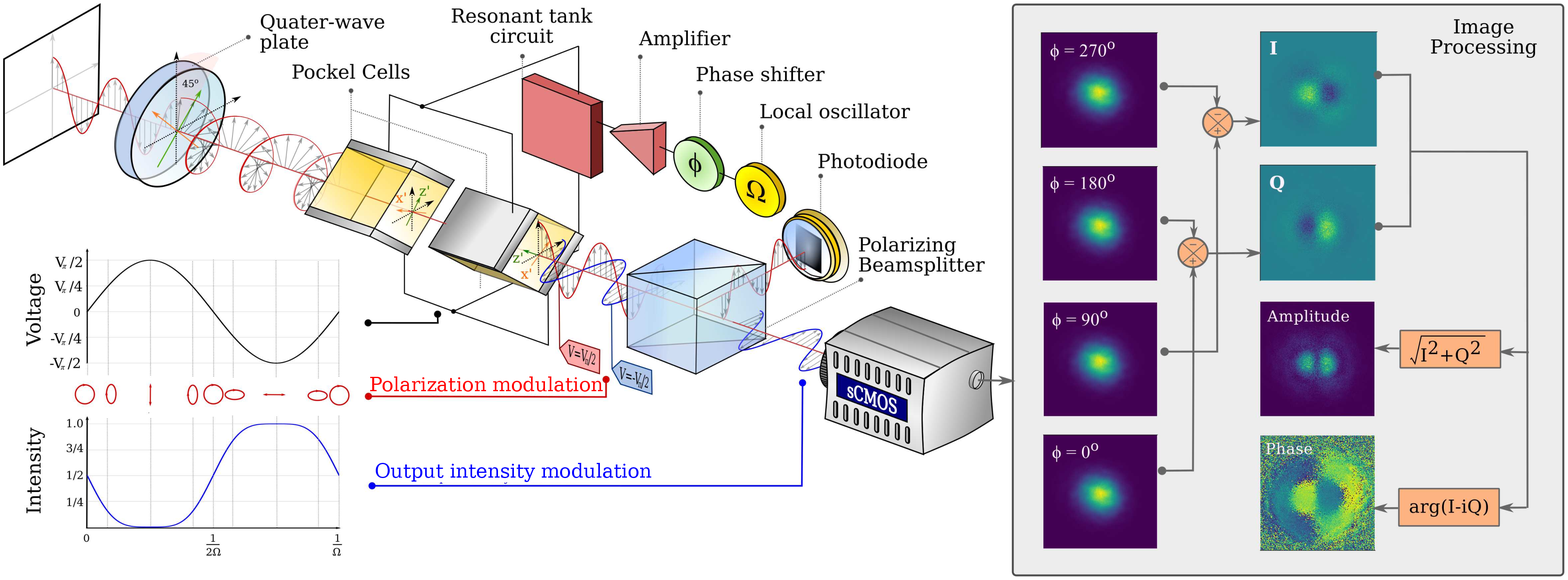}
	\caption{A schematic layout of the new camera. The quarter-wave plate, Pockels cell and polarizing beamsplitter form an optical switch that intensity modulates the beam incident on the sCMOS camera. Spatially-resolved magnitude and phase maps of the heterodyne beat between a reference field and a signal field that is frequency shifted from a reference field is calculated using four camera images acquired with modulation phases separated by $\pi/2$.}
	\label{fig:PhaseCamConcept_Layout}
\end{figure*}

To illustrate the operation of the new camera we consider a beam consisting of two components: a reference field $E_r(x,y) \exp[i(\omega_r t + \varphi_r(x,y))]$ and a signal field $E_s(x,y)\exp[i(\omega_s t + \varphi_s(x,y))]$. We wish to determine the spatial distribution of the amplitude and phase of the signal field relative to a reference field, which is phase-locked to and perhaps frequency offset from the input carrier field, as shown in magenta in Fig.~\ref{Fig:ifo_fields}.

Measuring this composite field using a photodetector would yield a voltage
\begin{equation}
\begin{split}
\ V(x,y) &\propto  E_r(x,y)^2 + E_s(x,y)^2 \\ 
& \hspace{0.5cm} + 2 E_r(x,y) E_s(x,y) \sin \left(\Omega t+ \varphi(x,y) \right)
\label{eq:voltage}
\end{split}
\end{equation}
where $\Omega = \omega_r -\omega_s$ and $\varphi(x,y)= \varphi_r(x,y) - \varphi_s(x,y)$.
However, the frequency of the heterodyne beat is much larger than the bandwidth of a typical pixelated camera and would not be measurable. Thus, we synchronously amplitude modulate the field incident on each pixel as shown in Fig.~\ref{Fig:WorkingPrinciple}. In this example, a square-wave amplitude modulation is applied to the beam at a frequency $\Omega$, with a phase $\phi=\varphi$ that yields the largest signal.
For in-phase modulation the pixel detector observes intensities that are greater than the unmodulated intensity, resulting in a DC output $(V_r + V_s)/2 + \delta V$, where the $V_{r/s}$ are due to the 
$E_{r/s}(x,y)^2$ terms in Eq.\ref{eq:voltage} and $\delta V$ is due to the RMS average of the $E_r(x,y) E_s(x,y)$ term. Similarly, for the modulation phase $\phi=\varphi + \pi$, the pixel observes intensities that are less than the unmodulated intensity, $(V_r + V_s)/2 - \delta V$. Subtraction of these provides $2\delta V \propto E_r(x,y) E_s(x,y)$~\cite{supl:1}.

The optimum demodulation phase $\phi$ is not known a priori. Thus we record four camera images, $V_\phi$ at $\phi = \{0, \pi/2, \pi, 3\pi/2\}$ for example. Combining these images yields the magnitude and phase of the heterodyne beat:
\begin{align}
\mathbf{I} &\equiv V_0 - V_\pi \\
\mathbf{Q} &\equiv V_{3\pi/2} - V_{\pi/2} \\
| E_r(x,y) E_s(x,y)| & \propto \sqrt{\left(\mathbf{I}\right)^2 + \left(\mathbf{Q}\right)^2 } ,\label{eq:amplitudeMap} \\
\varphi & = \arctan \left(-\frac{\mathbf{Q}}{\mathbf{I}}\right).\label{eq:phaseMap}
\end{align}
where we refer to $\mathbf{I}$ and $\mathbf{Q}$ as the "in-phase" and "quadrature" signals. The heterodyne beat has thus been demodulated to baseband by the optical switching, and hence the analogy to a lock-in amplifier.

A schematic of a practical realization is shown in Fig.\ref{fig:PhaseCamConcept_Layout}. The composite beam is first filtered using a polarizer and then circularly polarized using a quarter-wave plate. It then passes through a Pockels cell (PC) driven with a half-wave voltage that switches the polarization of the beam between \textit{s} and \textit{p} linear polarization. The polarizer converts this polarization modulation into an amplitude modulation. Typical camera images and the result of processing using Eq.~\ref{eq:amplitudeMap} and \ref{eq:phaseMap} are shown in Fig.~\ref{fig:PhaseCamConcept_Layout}.

The maximum image rate could in principle be doubled by recording both the transmitted and reflected beams simultaneously. In practice it is difficult to overlap the images from both cameras to enable an accurate subtraction. Additional differential effects, such as variation in the responsivity of the sCMOS arrays and aberrations in the polarizing beamsplitter, also reduce the performance in dual camera operation.

\section{Optical Lock-in cameras for gravitational wave detectors}\label{sec:gw}

The interferometer shown in Fig.\ref{Fig:ifo_fields} uses two sets of phase-modulation sidebands at 9~MHz and 45~MHz to control the length and alignment of the interferometer cavities~\cite{Izumi_2016}. The reflected RF fields are used to control the positions of the mirrors so that (a) the carrier is resonant in the the power recycling cavity (PRC) and arm cavities, (b) the 9~MHz sidebands are resonant in the PRC, and (c) the 45~MHz sidebands transmit through the PRC and are resonant in the SRC. Ideally, the upper and lower sidebands within each pair have the same spatial distribution and amplitude. However, as discussed earlier, differential wavefront distortion upsets this balance and degrades this ideal resonant condition.

Locations for phase cameras that could be used to investigate the sideband fields are also shown in Fig.\ref{Fig:ifo_fields}. In the simplest operating mode, a phase camera would analyze the heterodyne beat of the sampled carrier and a sideband field. An independent frequency-offset reference field could be used to diagnose the carrier and sideband fields individually. Imaging these simultaneously would require additional optical lock-in cameras. Alternatively the fields could be imaged sequentially with one camera which has a fixed demodulation frequency and its own frequency shifter. The frequency of the reference can then be changed to pick which RF field is demodulated and imaged.

The balance of the 9~MHz sideband pair and the mode-matching into the PRC can be analyzed using the \textit{Pick-off camera} and \textit{Reflection camera}, respectively. The balance of the 45~MHz sideband pair could be analyzed using the \textit{Anti-symmetric camera}. Additionally, the differential wavefront distortion leads to 9~MHz sideband fields in the SRC, resulting in a 36~MHz heterodyne beat.

The high spatial resolution and sampling speed of the optical lock-in camera could thus be used to measure the spatial distribution and amplitudes of individual sideband fields. These images can then be used to investigate the effect of differential wavefront distortion on the interferometer control, optimize thermal compensation systems, and investigate the effect of any high-spatial-frequency wavefront distortions.
 
Lastly, the field circulating within each arm cavity could be analyzed using the \textit{X and Y transmission cameras}. This will enable imaging of unexpected higher-order mode content in the arm cavities. For example, parametric instabilities which produce sidebands at $\approx$10--100~kHz around the carrier. The optical lock-in camera can image these fields and form part of future active control schemes to identify and suppress such instabilities~\cite{PhysRevD.91.092001, Ma_2017}.

\section{Test setup}\label{sec:expt}

\begin{figure}[!t]
\centering
\includegraphics[width=\columnwidth]{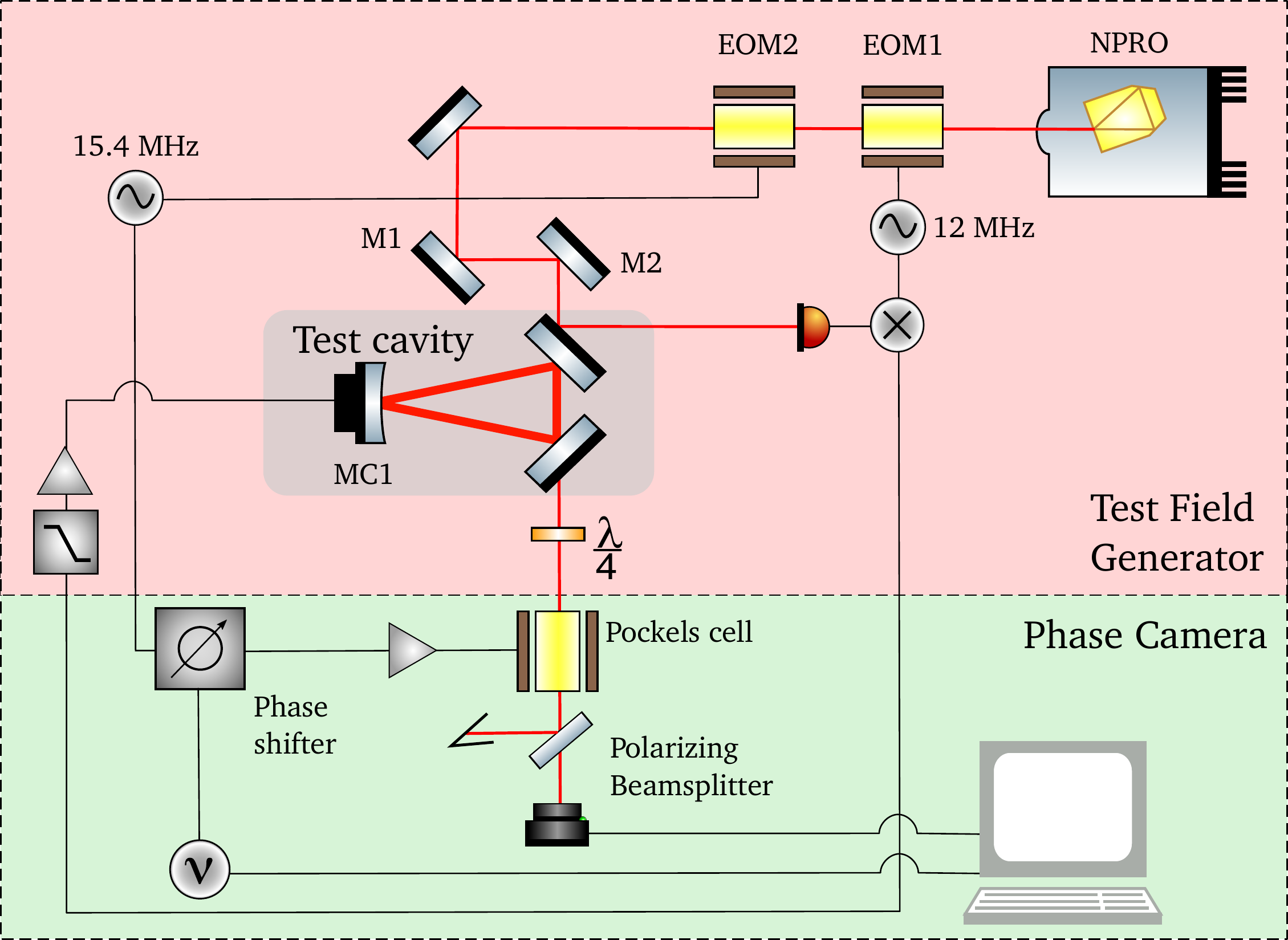}
\caption{Schematic of the optical system used to demonstrate the camera. The test field generator shown in the red box is used to produce a beam consisting of a reference and signal field.}
\label{fig:PhaseCamExperiment_Layout}
\end{figure}

We follow the approach used by~\citet{Goda:04} to demonstrate the operation and sensitivity of the optical lock-in camera. A schematic of the test system is shown in Fig.\ref{fig:PhaseCamExperiment_Layout}. It consists of two parts: a test field generator that produces a reference and signal field and the lock-in camera itself to image them. 

The test field consists of a large amplitude, TEM$_{00}$ mode and a higher-order mode of a high-finesse, $\approx 4000$, ring cavity that has a free spectral range of 540~MHz. The TEM$_{00}$ mode is produced by phase-locking a Nd:YAG NPRO to a TEM$_{00}$ mode of the ring cavity using the Pound-Drever-Hall technique~\cite{Black:2001} and the electro-optic phase modulator EOM1.

Higher-order modes are excited in the cavity by misaligning the incident beam using M1 and M2 and phase-modulating the beam at the cavity offset frequency using EOM2. The odd number of mirrors in the ring cavity breaks the resonance degeneracy between odd- and even-parity optical modes due to the odd-parity modes accumulating an additional $\pi$ phase shift during each round trip \cite{Waldman:2009,Arai:2013}. In our cavity, the TEM$_{30}$ and TEM$_{12}$ Hermite-Gauss modes resonate closest to the TEM$_{00}$ mode, at offset frequencies of 15.7~MHz and 15.3~MHz respectively.

\begin{figure}[ht]
\centering
\includegraphics[width=0.4\textwidth]{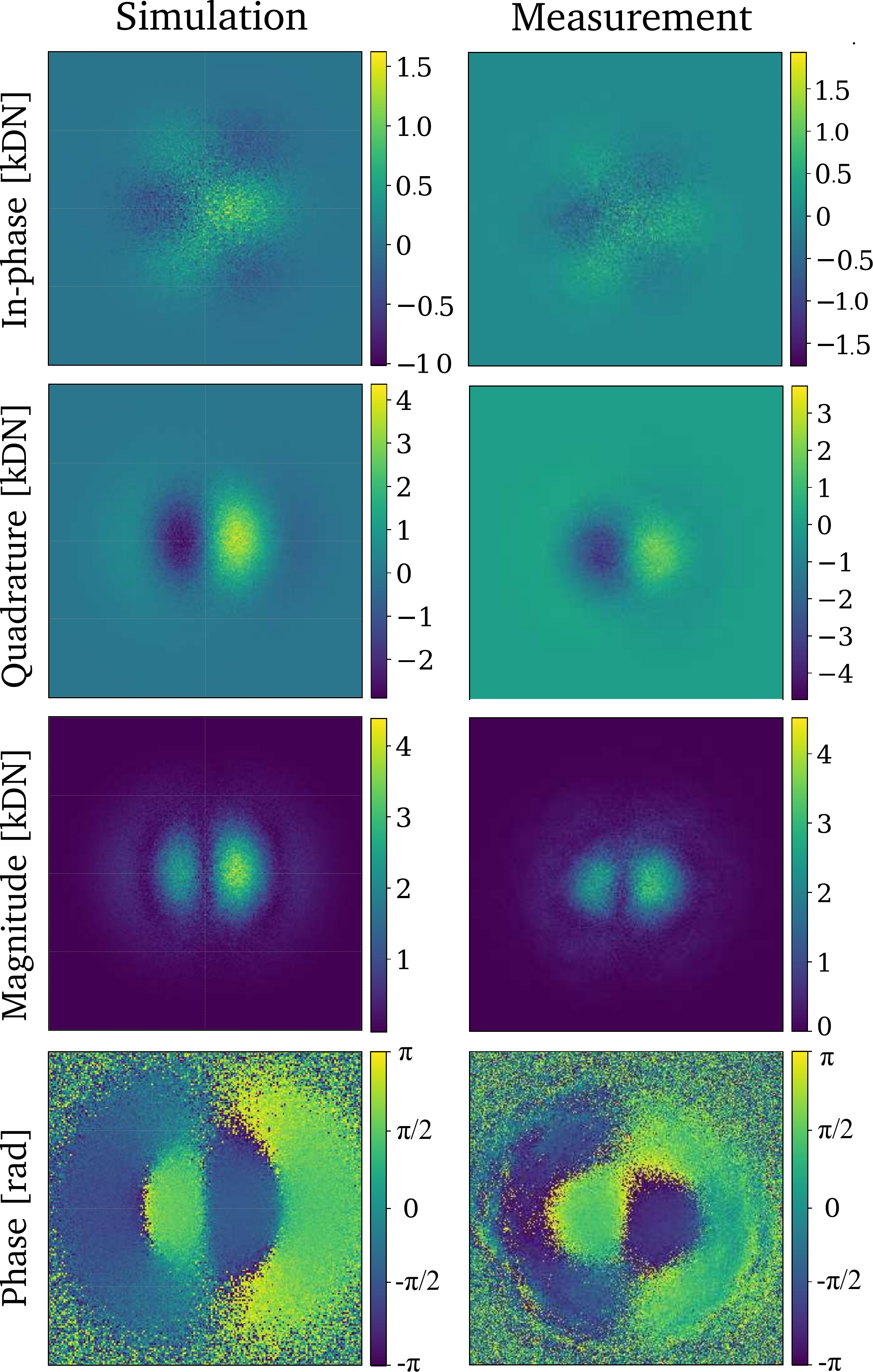}
\caption{Comparison between camera measurements and the predictions from a \textsc{Finesse} simulation. The digitized pixel values are given in units of thousands of digital-numbers (kDN) and plotted using the false-color scale bars.}
\label{Fig:Result_singleframe}
\end{figure}

For the test described here, we chose to drive EOM2 at 15.4~MHz as it enabled the excitation of both modes. The beam emitted by the ring cavity therefore consists of a large-amplitude TEM$_{00}$ reference field with frequency $\omega_r$, and a smaller-amplitude TEM$_{30}$ and TEM$_{12}$ signal field oscillating mostly at the 15.4~MHz-shifted frequency, $\omega_s$. 

The performance of the camera is affected by the sCMOS properties. A high dynamic range, bit-depth, and linearity are crucial as we must subtract images to remove the offset due to the high power carrier. A high frame rate is also required as four frames are required to produce the intensity and phase images, and to allow averaging of of shot noise, provided it does not result in an unacceptable reduction in dynamic range or spatial resolution.

In this work we use a Zyla~4.2 sCMOS camera, which has a sensor size of 2048x2048 pixels, a dynamic range of 89~dB, a 16-bit readout, a maximum frame rate of 100~fps and a quantum efficiency of 5\% at 1064~nm. The camera window was anti-reflection coated for the 1064~nm. The rolling-frame shutter for this camera does not affect the measurement process as the demodulation phases for each pixel are still separated by $\pi/2$.

\section{Results}

\begin{figure*}[t]
\centering
\includegraphics[width =\textwidth]{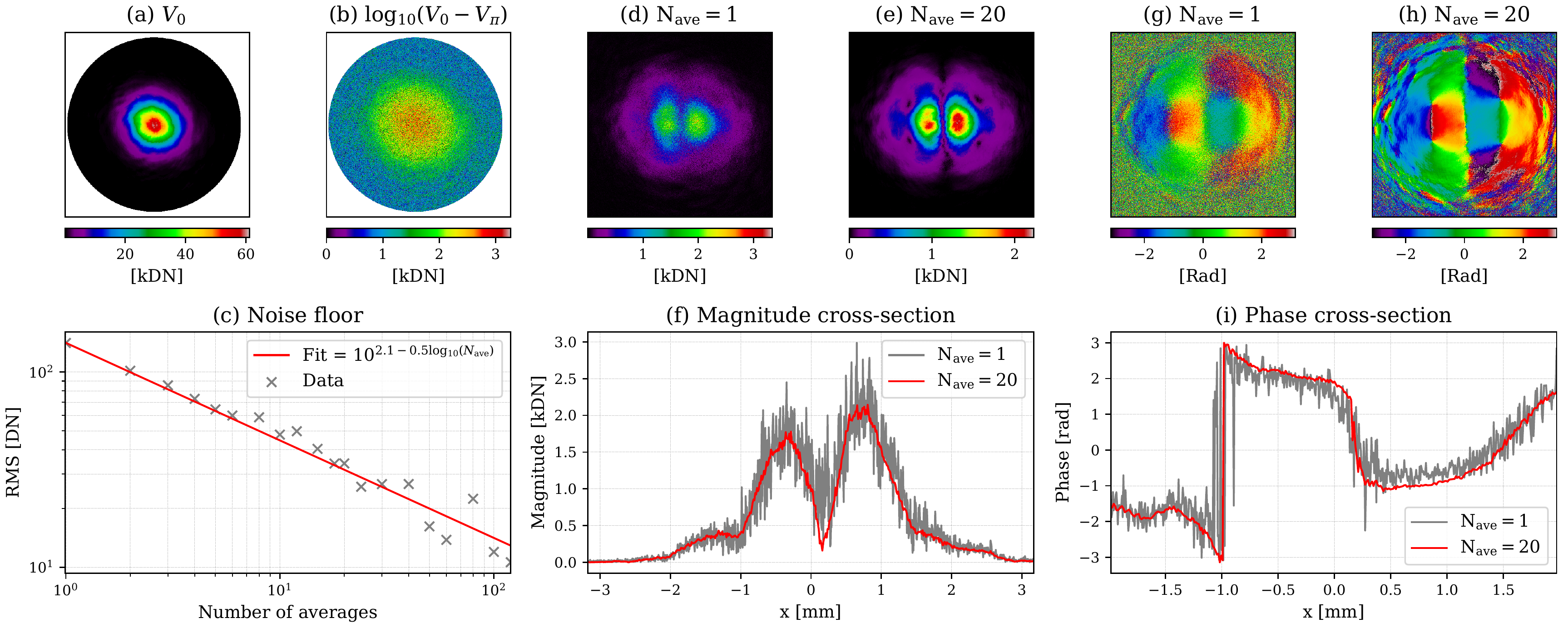}
\caption{Typical images of (a) $V_0$ or $V_\pi$ image, and (b) $\log_{10}(|V_0-V_\pi|)$ for a single pair of images. (c) Shows how the RMS of $|V_0-V_\pi|$ decreases with averaging. (d, e) Maps of the magnitude of the heterodyne beat for $N_\mathrm{ave}=1$ and $N_\mathrm{ave}=20$. (g,h) Maps of the phase of the heterodyne beat for $N_\mathrm{ave}=1$ and $N_\mathrm{ave}=20$.
Images (e) and (h) were taken with $2\times2$ pixel binning. (f) Plot of the magnitude variation along the center of (d) and (e). (i) Plot of the Phase variation along the center of (f) and (g).}
\label{Fig:Noisefloor}
\end{figure*}
	
Typical $\mathbf{I}$ and $\mathbf{Q}$ images and the result of a numerical simulation of the test-field generator using \textsc{Finesse}\cite{Finesse} are shown in Fig.~\ref{Fig:Result_singleframe}. In this case, the TEM$_{30}$ mode is apparent in the Q demodulation while the TEM$_{12}$ mode occurs mostly in the I demodulation. Only the two central maxima of the TEM$_{30}$ mode are observed in this demonstration as the amplitude of the TEM$_{00}$ reference field is much smaller at the location of the outer maxima. 

The \textsc{Finesse} simulation used plausible misalignments and included shot noise to reproduce outputs of the optical system.  For the simulation shown in Fig.\ref{Fig:Result_singleframe}, the ratio of the power in higher-order mode to that in the TEM$_{00}$ was 14\% for the TEM$_{30}$ and 8\% for the TEM$_{12}$ modes, and thus the magnitude is dominated by the TEM$_{30}$ mode but the phase shows some influence of the weaker TEM$_{12}$ mode, which degrades the spatial resolution we are able to demonstrate below.
	
The sensitivity of the optical lock-in camera was investigated by removing the 15.4~MHz modulation from EOM2 and recording frames with the demodulation phase alternating between 0 and $\pi$. An image typical of individual $V_0$ and $V_\pi$ frames is shown in Fig.~\ref{Fig:Noisefloor}(a). The magnitude of a typical $V_0-V_\pi$ image is show in Fig.\ref{Fig:Noisefloor}(b). The $RMS$ average of the residual values can be decreased by averaging multiple $V_0-V_\pi$ pairs as shown in Fig.\ref{Fig:Noisefloor}(c). It is also apparent from Fig.\ref{Fig:Noisefloor}(c) that the decrease in the $RMS$ is $\propto 1/\sqrt{N_\text{ave}}$ where $N_\text{ave}$ is the number of pairs in the average, thereby showing that the residuals in Fig.\ref{Fig:Noisefloor}(b) are due to pixel shot noise.

The improvement in sensitivity due to averaging was demonstrated by reinstating the 15.4~MHz modulation of EOM2 and recording twenty frames at each of the four demodulation phases. The magnitude and phase of the beat with $N_\text{ave}=1$ and $N_\text{ave}=20$ are shown in Fig.~\ref{Fig:Noisefloor}(d) and (e), and (f) and (g) respectively. Averaging over 20 frames improves the signal-to-noise ratio in the maps as seen in Fig.~\ref{Fig:Noisefloor}(f) and (i). In addition to the averaging, pixel-binning can also be employed for further SNR improvements without sacrificing speed---as was used for the $N_\text{ave}=20$ cases above, where $2\times2$ binning was employed.

	\begin{figure*}[t]
		\centering
		\includegraphics[width=\textwidth]{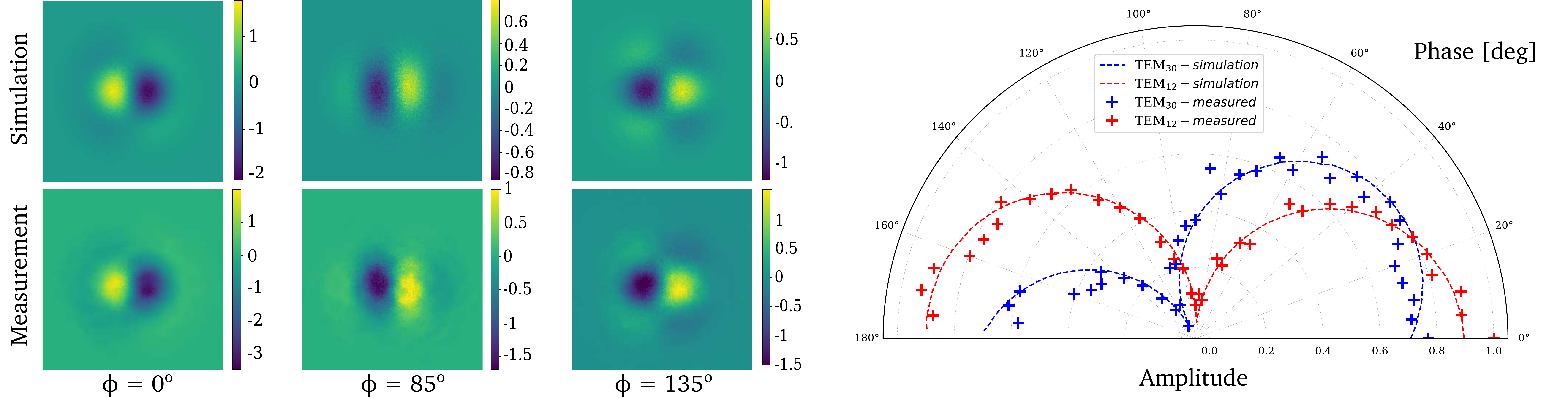}
		\caption{The measured and simulated demodulated signal mode content.  $\phi=0^o, 85^o, 135^o$ are shown on the left with the corresponding simulation showing the individual modes. The data and model have been scale normalized.}
		\label{Fig:PhaseCam_sweep}
	\end{figure*}

The minimum signal power detectable can be estimated from the ratio of the digital number~(DN) noise on the central peaks in Fig.~\ref{Fig:Noisefloor}(g), approximately 0.1~kDN, to the DN of the reference field in Fig.~\ref{Fig:Noisefloor}(a), approximately 60~kDN: as $2E_sE_r/(E_r)^2 \approx 0.1/60$ and thus $(E_s/E_r)^2 \approx -62$~dB below the power in the reference field, a 12 dB improvement on that reported by~\citet{Goda:04}.

The relatively poor signal-to-noise associated with the outer maxima of the TEM$_{30}$ signal field is due to the small diameter of the TEM$_{00}$ reference field in the test system.  It could be improved by using a larger diameter reference field that is frequency-offset locked to the signal field, or by using a liquid crystal attenuator or spatial light modulator~\cite{Nayar2003,ZHONGDONG201431,MANNAMI2007359}.

To analyze the output of phase cameras it will be important to extract the relative phase of the higher order modes in a beam. Fig.\ref{Fig:PhaseCam_sweep} shows how the modal content extracted from the in-phase and quadrature images varies with demodulation phase. We can see that the TEM$_{12}$ mode is out-of-phase with the carrier at 85$^\circ$ and the TEM$_{30}$ at 135$^\circ$---this phase relationship agrees well with that predicted by the \textsc{Finesse} model.

\section{Conclusion}

In this work we have introduced a new type of phase camera, the optical lock-in camera, and demonstrated its ability to produce high spatial resolution maps of the phase and intensity of a coherent light field. This is achieved with a higher acquisition rate and resolution than previous phase camera implementations. The camera is both more compact and does not rely on any mechanically moving parts, thus reducing scattered light and enabling operation during scientific observations in gravitational wave interferometers.

The phase and intensity of a specific frequency component of a beam is imaged by creating a heterodyne beat with a reference field and synchronously amplitude modulating it. The key element is the Pockels cell which acts as a fast optical switch to provide the amplitude modulation. By switching over the entire field optically, rather than electronically, and imaging with a sCMOS array, each pixel can behave as an optical lock-in amplifier.

The results of our proof-of-principle measurements are in excellent agreement with the predictions of a theoretical \textsc{Finesse} model in our test system. We also demonstrate that the sensitivity is limited purely by shot-noise and can be improved by simple averaging, resulting in a noise floor of -62~dBc from data recorded in 2s. The performance can be easily improved by using faster or more sensitive cameras, such as InGaAs arrays which can achieve $>100$~Hz frame rates, or by sacrificing spatial resolution for faster acquisition rates on dense sCMOS arrays, by region-of-interest extraction or pixel-binning. 

Various applications of this camera in advanced gravitational wave detectors have been highlighted. The additional information provided by them should enable better diagnostics of high spatial frequency effects within an interferometer. This will provide a new tool for improving both their duty-cycle and sensitivity. This will be particularly important for the thermal compensation systems as ever increasing stored optical power is used in current and future generations of detectors. These cameras can also offer the ability to image physical processes such as parametric instabilities, offering a new method to monitor them or to act as a sensor in an active suppression scheme. 

This project was funded by the Australian Research Council grants DP150103359 and CE170100004.

\bibliographystyle{unsrtnat}
\bibliography{PhaseCameraBib}

\clearpage

\end{document}